\documentclass[acmsmall, authorversion, nonacm, natbib=false]{acmart}

\usepackage[ruled]{algorithm2e}  % Algorithm environment (https://www.ctan.org/pkg/algorithm2e)
\usepackage{amsthm}  % Required for various environments (https://www.ctan.org/pkg/amsthm)
    \theoremstyle{remark}
        \newtheorem{remark}{Remark}
            
\usepackage[datamodel=acmdatamodel, style=acmnumeric]{biblatex}  % Sophisticated bibliographies (https://ctan.org/pkg/biblatex)
    \patchcmd{\bibsetup}{\interlinepenalty=5000}{\interlinepenalty=10000}{}{}
    \DeclareFieldFormat{titlecase:title}{\MakeSentenceCase*{#1}}
    \DeclareSourcemap{
        \maps[datatype=bibtex]{
            \map{
                \step[fieldsource=doi, final]
                \step[fieldset=url, null]
                \step[fieldset=isbn, null]
            }
            \map{
                \step[fieldsource=doi, match=\regexp{\\_}, replace=\regexp{_}]
                \step[fieldsource=url, match=\regexp{\\_}, replace=\regexp{_}]
            }
            \map{
                \step[fieldsource=eprinttype, final]
                \step[fieldset=doi, null]
                \step[fieldset=url, null]
                \step[typesource=article, typetarget=online]
            }
            \map{\step[fieldsource=url, match=\regexp{eprint\.iacr\.org}, replace=\regexp{ia\.cr}]}
            \map{
                \step[fieldset=address, null]
                \step[fieldset=location, null]
                \step[fieldset=editor, null]
                \step[fieldset=publisher, null]
            }
        }
    }
\usepackage{csquotes}  % Context-aware quotes (https://www.ctan.org/pkg/csquotes)
\usepackage{graphicx}  % Figures and .eps files (https://www.ctan.org/pkg/graphicx)
    \newsavebox{\imagebox}  % Allow vertical alignment of subfigures independent of caption alginment
\usepackage{mathtools}  % Custom operators (https://www.ctan.org/pkg/mathtools)
    \DeclareMathOperator{\floor}{floor}
    \DeclareMathOperator{\girth}{girth}
    
    \DeclareMathOperator{\neigh}{neigh}
    \DeclarePairedDelimiter{\abs}{\lvert}{\rvert}
    \DeclarePairedDelimiter{\range}{\lBrack}{\rBrack}
    \DeclarePairedDelimiter{\norm}{\lVert}{\rVert}
\usepackage{multirow}  % Span multiple rows and columns (and make alignment easier) (https://www.ctan.org/pkg/multirow)
\usepackage{pgfplots}  % Parsing figures exported with matlab2tikz (https://www.ctan.org/pkg/pgf)
    \pgfplotsset{compat=newest}  % Recommended by matlab2tikz
    \pgfplotsset{plot coordinates/math parser=false}  % Recommended by matlab2tikz
\usepackage{siunitx}  % Comprehensive SI units  (https://ctan.org/pkg/siunitx)
\usepackage{subcaption}  % Subfigures (https://ctan.owrg/pkg/subcaption)
\usepackage{tabularx}  % Width-aware tables (https://ctan.owrg/pkg/tabularx)
    \newcolumntype{C}{>{\centering\arraybackslash}X}  % Max-width column
\newcommand*{\tikzsetnextfilename}[1]{}
\usepackage{xcolor}  % Named and custom colours (https://ctan.owrg/pkg/xcolor)
    \definecolor{matlab1}{HTML}{0072BD}  % Blue
    \definecolor{matlab2}{HTML}{D95319}  % Orange
    \definecolor{matlab3}{HTML}{EDB120}  % Yellow
    \definecolor{matlab4}{HTML}{7F2F8E}  % Purple (violet)
    \definecolor{matlab5}{HTML}{77AC30}  % Green
    \definecolor{matlab6}{HTML}{4DBEEE}  % Cyan
    \definecolor{matlab7}{HTML}{A2142F}  % Burgundy
\setcopyright{rightsretained}
\copyrightyear{2025}

\begin{document}
    \title{Optimal Graph Stretching for Distributed Averaging}

    \author{Florine W.\ Dekker}
    \orcid{0000-0002-0506-7365}
    \affiliation{%
        \institution{Delft University of Technology}
        \city{Delft}
        \country{Netherlands}}
    \email{f.w.dekker@tudelft.nl}

    \author{Zekeriya Erkin}
    \orcid{0000-0001-8932-4703}
    \affiliation{%
        \institution{Delft University of Technology}
        \city{Delft}
        \country{Netherlands}}
    \email{z.erkin@tudelft.nl}

    \author{Mauro Conti}
    \orcid{0000-0002-3612-1934}
    \affiliation{%
        \institution{Università di Padova}
        \city{Padua}
        \country{Italy}}
    \affiliation{%
        \institution{Delft University of Technology}
        \city{Delft}
        \country{Netherlands}}
    \email{mauro.conti@unipd.it}

    \begin{abstract}
    The performance of distributed averaging depends heavily on the underlying topology.
    In various fields, including compressed sensing, multi-party computation, and abstract graph theory, graphs may be expected to be free of short cycles, i.e.\ to have high girth.
    Though extensive analyses and heuristics exist for optimising the performance of distributed averaging in general networks, these studies do not consider girth.
    As such, it is not clear what happens to convergence time when a graph is stretched to a higher girth.

    In this work, we introduce the \emph{optimal graph stretching problem}, wherein we are interested in finding the set of edges for a particular graph that ensures optimal convergence time under constraint of a minimal girth.
    We compare various methods for choosing which edges to remove, and use various convergence heuristics to speed up the searching process.
    We generate many graphs with varying parameters, stretch and optimise them, and measure the duration of distributed averaging.
    We find that stretching by itself significantly increases convergence time.
    This decrease can be counteracted with a subsequent repair phase, guided by a convergence time heuristic.
    Existing heuristics are capable, but may be suboptimal.
\end{abstract}



    \maketitle

    \section{Introduction}\label{sec:introduction}
Distributed averaging allows nodes in a peer-to-peer network to find the global mean of the nodes' local values in a completely distributed manner.
Throughout the protocol's iterative process, each node's estimate of the global mean continues to improve until a consensus is reached.
Distributed averaging has applications in various fields, including gossip learning~\cite{BoydGPS05}, fully-distributed learning~\cite{VanhaesebrouckB17}, and control systems~\cite{HadjicostisDC18}.
In all cases, the challenge is to find an algorithm that is efficient in terms of convergence time and communication cost.

The study of convergence in consensus algorithms is heavily tied to studies on \emph{synchronisability} in chaos theory, which, roughly speaking, studies the ability of disjoint systems to synchronise spontaneously~\cite{PecoraC98, BarahonaP02}.
We know from chaos theory that the convergence time of distributed averaging is heavily tied to the underlying topology~\cite{BoydGPS05, LiDCH10}.
Optimising a topology for convergence time is hard~\cite{XiaoB04}, and so a multitude of heuristics have been proposed, including those based on graph metrics such as degree, closeness centrality, and efficiency~\cite{HagbergS08, SirocchiB22}, and on spectral metrics such as eigenratio and algebraic connectivity~\cite{GhoshB06, XiaoB04}.

Meanwhile, several fields study the girth of the network, which is the length of its shortest cycle.
In compressed sensing, high girth positively impacts reconstruction guarantees~\cite{KhajehnejadTDH11, LiuX13}.
In multi-party computation, the girth implies specific privacy guarantees~\cite{DekkerEC25a}.
Finally, in graph theory, high-girth graphs are an interesting concept per se~\cite{Margulis82}, and are important when studying expander graphs~\cite{Paredes21}.
Various authors have also proposed algorithms for increasing the girth of an existing graph.
Algorithms for coding theory focus on bipartite graphs~\cite{HuEA05, LauTT11}, while algorithms for expander graphs focus on degree-regular graphs~\cite{Paredes21}.

To the best of our knowledge, there are no works that study the intersection of these two areas.
Therefore, in this work, we ask:
How does \enquote{stretching} the girth of a graph to a higher value affect the convergence time of distributed averaging?
Additionally, we ask how to minimise the number of leaf nodes, since these are undesirable in various applications~\cite{AlonHL02, DekkerEC25a}.
To answer both our questions, we formalise our optimisation problem, consider several stretching and leaf minimisation algorithms, optimisation heuristics, and graph families, and compare the results.

We find that stretching a graph to a higher girth significantly increases the convergence time, typically by an order of magnitude.
Since stretching consists solely of removing edges, we find that the best algorithm prioritises the removal of those edges that are in the largest number of cycles.
Additionally, lost convergence time can be recuperated partially by greedily optimising the edge set using a heuristic for convergence time.
Meanwhile, minimising the number of leaves has little impact on convergence time, with little difference between the various algorithms studied.
Finally, though the studied heuristics are adequate for improving convergence time, our results indicate that heuristics tailored for high-girth graphs may be able to achieve even better convergence time.

In \autoref{sec:preliminaries}, we present our notation and various preliminaries.
In \autoref{sec:related-work}, we survey related work.
In \autoref{sec:optimal-graph-stretching-problem}, we introduce the optimal graph stretching problem and our exact research questions.
In \autoref{sec:method}, we explain our research method.
In \autoref{sec:results}, we present our results.
Finally, in \autoref{sec:conclusion}, we offer our conclusions.

    \section{Preliminaries}\label{sec:preliminaries}
In general,
we denote the first element of a vector~$v$ by~$v_0$,
the absolute value of a scalar~$x$ by~$\abs{x}$,
the number of elements in a collection~$S$ by~$\abs{S}$,
the range of integers~$\{ 0 \ldots n - 1 \}$ by~$\range{n}$,
and the Euclidian norm of a vector~$v$ by~$\norm{v}_2$.

\subsection{Graph theory}\label{subsec:preliminaries:graph-theory}
\paragraph{Basics}
A graph~$G = (V, E)$ is a set of vertices~$V$ and a set of edges~$E \subseteq V \times V$.
In this work, we consider only simple graphs, i.e.\ unweighted, undirected, self-loopless graphs, where each edge may occur at most once.
For any node~$v \in V$, the function~$\neigh(v)$ gives the set of direct neighbours of~$v$, and~$\deg(v)$ gives the degree of~$v$.
The adjacency matrix~$A$ of graph~$G$ is a $\abs{V}$-by-$\abs{V}$-matrix where, for any~$i, j \in \range{\abs{V}}$, we have~$A_{i, j} = 1$ if~$(V_i, V_j) \in E$ and~$A_{i, j} = 0$ otherwise.
The (unoriented) incidence matrix~$B$ of graph~$G$ is a $\abs{V}$-by-$\abs{E}$-matrix where, for any~$i \in \range{\abs{V}}, j \in \range{\abs{E}}$, we have~$B_{i, j} = 1$ if~$V_i \in E_j$ and~$B_{i, j} = 0$ otherwise.

\paragraph{Spectral theory}
For any $n$-by-$n$ matrix $M$, an eigenvector $v$ is a vector such that $Mv = \lambda v$ for some scalar $\lambda$.
This scalar $\lambda$ is the eigenvalue corresponding to $v$.
The matrix $M$ has $n$ (not necessarily unique) eigenvalues, collectively known as the \textit{spectrum} of $M$.
For any $1 \leq i \leq n$, we write $\lambda_i(M)$ to mean the $i$th-smallest eigenvalue of $M$.
That is, $\lambda_1(M) \leq \lambda_2(M) \leq \ldots \leq \lambda_n(M)$.
We drop the index $M$ when the matrix is clear from context.

\paragraph{Spectral graph theory}
The Laplacian~$L$ of a graph~$G$ is the $\abs{V}$-by-$\abs{V}$ matrix~$BB^T$.
For any~$i, j \in \range{\abs{V}}$, we have~$L_{i, j} = -A_{i, j}$ if $i \neq j$ and~$L_{i, j} = \deg(V_i)$ otherwise.
Some eigenvalues of~$L$ are special:
$\lambda_1 = 0$;
$\lambda_2$ is called the algebraic connectivity (and the associated eigenvector is called the Fiedler vector);
$\lambda_n$ is called the spectral radius;
and $\frac{\lambda_2}{\lambda_n}$ is called the eigenratio.
The algebraic connectivity $\lambda_2 = 0$ if and only if $G$ is connected~\cite{Fiedler73}.
All eigenvalues increase monotonically with the edge set.
(This cannot be said for the eigenratio.)
Formally, given graphs~$G_1 = (V, E_1)$ and~$G_2 = (V, E_2)$ where~$E_1 \subseteq E_2$, we have~$\lambda_i(L_1) \leq \lambda_i(L_2)$~\cite{Fiedler73}.
In fact, the eigenvalues of the two graphs become interlaced~\cite{GroneMS90, Russell91}:
$\lambda_i(L_1) \leq \lambda_i(L_2) \leq \lambda_{i + 1}(L_1) \leq \lambda_{i + 1}(L_2)$.

\subsection{Distributed averaging}\label{subsec:preliminaries:distributed-averaging}
Consider a graph~$G = (V, E)$ with $n \coloneqq \abs{V}$~nodes.
Each node~$v \in V$ has a scalar value~$x_v$ and can communicate only with their direct neighbours~$\neigh(v)$.
In distributed averaging, the task for each node is to find the global mean~$\frac{\sum_{v \in V} x_v}{n}$.

Distributed averaging can be achieved using a distributed asynchronous push-pull algorithm:
Nodes iteratively calculate the mean of their local neighbourhood and then replace their own value with that mean.
Specifically, in this work, the algorithm we consider has the following properties:
\begin{itemize}
    \item
    \emph{Asynchronous~\cite{BoydGPS06}:}
    Users do not coordinate to choose which user is next.
    Instead, users randomly and independently \enquote{wake up} and perform their iteration.

    \item
    \emph{Linear iterations~\cite{Olfati-SaberM03, XiaoB04}:}
    Distributed averaging algorithms differ in which neighbours are included in the averaging operation.
    To achieve convergence, it is sufficient that each direct neighbour is selected with a non-zero probability~\cite{HadjicostisDC18}.  % See Section 3.3.2
    For simplicity, in our implementation, the initiating user selects one of its neighbours at random.

    \item
    \emph{Push-pull~\cite{DemersGHILSSST88}:}
    The mean calculated by the initiating user is used as the new local value of both the initiating user~$v$ (\enquote{pull}) and the selected neighbour~$w$ (\enquote{push}).
\end{itemize}

Implementing this type of distributed averaging requires each user to simultaneously run two threads: one to initiate rounds, and one to respond.
We show the corresponding algorithms respectively in \autoref{alg:distributed-averaging:active} and \autoref{alg:distributed-averaging:passive}.
To avoid overly complex notation, these algorithms do not address issues relating to concurrency.

\begin{minipage}[t]{.45\linewidth}
    \vspace{0pt}
    \begin{algorithm}[H]
        \caption{Active thread of each user~$v$ in distributed averaging}
        \label{alg:distributed-averaging:active}
        \While{true}{
            \FuncSty{sleep()}\;
            $w \gets_R \neigh(v)$%
            \tcp*[r]{random sample}
            send $x_v$ to $w$\;
            receive $x_w$ from $w$\;
            $x_v \gets \frac{x_v + x_w}{2}$\;
        }
    \end{algorithm}
\end{minipage}%
\hfil%
\begin{minipage}[t]{.45\linewidth}
    \vspace{0pt}
    \begin{algorithm}[H]
        \caption{Passive thread of each user~$w$ in distributed averaging}
        \label{alg:distributed-averaging:passive}
        \While{true}{
            receive $x_v$ from $v$\;
            send $x_w$ to $v$\;
            $x_w \gets \frac{x_w + x_v}{2}$\;
        }
    \end{algorithm}
\end{minipage}

    \section{Related work}\label{sec:related-work}
To the best of our knowledge, there is no literature that covers the relation between distributed algorithm convergence speed and graph girth.
Therefore, in this section, we survey those works that are most closely related.
In \autoref{subsec:related-work:convergence}, we discuss works on the relation between topology and convergence.
In \autoref{subsec:related-work:girth}, we discuss works on high-girth graphs and short-cycle removal.

\subsection{Convergence}\label{subsec:related-work:convergence}
There exists a vast body of work that analyses the relation between topology and convergence.
These works have their origin in physics, aiming to predict the ability of dynamic networks to spontaneously synchronise~\cite{PecoraC98, BarahonaP02}.
Since similar dynamics occur in distributed systems, results on synchronisability were adopted into computer science, where the concept is referred to as convergence~\cite{DonettiHPMI05, LiFXZ11, LiuYTL14}.
For simplicity, in the following overview, we will speak of convergence even if the cited work is about synchronisation.

\paragraph{Spectral theory}
% Synchronisability
\textcite{PecoraC98} and \textcite{BarahonaP02} show that the convergence speed of a graph is determined by the eigenvalues of that graph's Laplacian.
Subsequent literature often uses algebraic connectivity and eigenratio as heuristics of the graph's convergence speed.

% Distributed consensus
\textcite{KarAM06} show that (non-bipartite) Ramanujan graphs exhibit high convergence speeds, both as expected from their eigenratio, and as validated in numerical simulations.
The authors point to various constructions of Ramanujan graphs in literature.

% Synchronisability = distributed consensus
\textcite{DonettiHPMI05} propose a new family of graphs that achieve fast convergence: entangled networks.
They propose an algorithm that finds entangled networks with a desired number of nodes and average degree.
The algorithm starts with an arbitrary graph and, in each iteration, chooses random pairs of edges, performs an edge exchange on each edge pair~$\left((e_1, e_2), (e_3, e_4)\right)$ to get~$\left((e_1, e_4), (e_2, e_3)\right)$, and accepts the change if the eigenratio decreases.
By using simulated annealing, the algorithm avoids getting stuck in local optima.
\textcite{DonettiNM06} extend their analysis, and show that entangled networks correspond exactly to so-called cage graphs and Ramanujan graphs.
However, the authors conclude that the aforementioned algorithm is inefficient for finding Ramanujan graphs compared to existing literature.

% Synchronisability
\textcite{WangZXK07} improve upon the aforementioned edge exchange algorithm by using tabu search instead of simulated annealing.
The authors also observe that the clustering coefficient is a good heuristic to predict convergence speed, and show that basing the search algorithm's acceptance criterion on the clustering coefficient also creates graphs with high convergence speeds.

% Independent of synchronisability vs. convergence
\textcite{GhoshB06} propose a greedy algorithm to optimise algebraic connectivity.
At each iteration, find the Fiedler vector~$u$, and add the edge~$(i, j)$ with largest~$(u_i - u_j)^2$.
Since the work focuses on optimising algebraic connectivity, it is not clear how this algorithm affects the convergence speed of distributed averaging.

\paragraph{Degree relations}
% Synchronisability
\textcite{RadJH08} propose an algorithm that removes edges based on the sum of adjacent node degrees, and adds edges using the Fiedler vector criterion of \textcite{GhoshB06}, and shows that this results in a network with optimised eigenratio, which coincides with Ramanujan graphs.
The authors note that many other metrics provide similar results.

% Synchronisability, and later convergence
In a series of works, \textcite{YangT11}, \textcite{YeungYTL12}, and \textcite{LiuYTL14} create increasingly performant heuristics for maximising convergence speed.
Ultimately, they settle on a tabu search-based algorithm in which edges are removed and added as done by \textcite{RadJH08}, and accept the resulting candidates depending on whether the eigenratio improved.
The algorithm prefers adding edges between nodes that are within a short distance of each other in the underlying physical network, and ensures that the resulting graph is connected.

% Synchronisability
However, \textcite{DonettiHM08} show that while degree-degree associations of neighbouring nodes indeed correlate negatively with the network's convergence speed, this correlation is not causative, as the mere act of introducing such heterogeneity does not by itself decrease the eigenratio.

\paragraph{Comparisons}
% Synchronisability
\textcite{HagbergS08} compare a multitude of greedy edge-modifying algorithms to determine which methods achieve convergence in the fewest iterations.
Overall, they conclude that methods that focus on increasing algebraic connectivity outperform those based on spectral radius and degree criteria, and that edge exchanges are not necessarily better than separate edge additions and removals.
The authors do not consider eigenratio as a separate optimisation metric.
% Finally, \textcite{MahdiA09} comment that this algorithm gets stuck in local minima, which can be avoided using simulated annealing instead of greedy search.

% Convergence
\textcite{SirocchiB22} extensively compare metrics and find that the metrics that most strongly correlate with high convergence speed of a distributed consensus protocol are high closeness centrality, implying that information travels quickly, and small clustering coefficient, implying that information is sent non-redundantly.
However, these metrics vary in their accuracy for different graph families.
Unfortunately, the authors do not investigate eigenratio as a metric.

\subsection{Girth}\label{subsec:related-work:girth}
We discuss works related to (increasing) girth in graphs.

\paragraph{Moore bound}
Firstly, we note the Moore bound~\cite{Biggs93}.
For $d$-regular graphs with girth~$g$, the number of nodes must be at least
\begin{equation}
    \label{eq:moore-bound}
    \begin{cases}
        2 \sum_{i = 0}^{g / 2 - 1} (d - 1)^i, & \text{if girth is even} \\
        1 + d \sum_{i = 0}^{(g - 1) / 2 - 1} (d - 1)^i, & \text{if girth is odd.}
    \end{cases}
\end{equation}
\textcite{AlonHL02} show that if $d$ is taken to be the graph's average degree, and each node has at least degree two, \autoref{eq:moore-bound} also holds for irregular graphs.
Consequently, another way to interpret the Moore bound is to say that, given the number of nodes and a desired girth, there is an upper limit on the number of edges.
Therefore, when a higher girth is desired, the Moore bound dictates that it may be necessary to remove some edges.

\paragraph{High-girth graph constructions}
We note several works that present algorithms for constructing graphs with high girth.
Though these works do not consider increasing girth in arbitrary existing graphs, the algorithms are interesting nonetheless.

\textcite{Chandran03} provides a construction of high-girth almost-regular graphs.
Briefly, this algorithm takes the number of nodes~$n$, the desired average degree~$k < \frac{n}{3}$, and outputs a graph with girth~$g \geq \log_k(n) + O(1)$.
The algorithm starts with $n$~nodes and the edges being a perfect matching on those nodes, and then iteratively adds edges between the most distant pair of nodes such that at least one of the nodes in the pair is a node with the lowest degree globally.
The graph is almost regular in the sense that any two nodes differ in degree by at most two.

\textcite{LinialS21} provide a construction of high-girth regular graphs.
Their procedure is similar to that of \textcite{Chandran03}, but starts with a Hamiltonian cycle~$G$ on $n$~vertices instead, and, with high probability, gives a $k$-regular graph with girth at least~$c \log_{k - 1}(n)$ for input $0 < c < 1$.

Finally, \textcite{LazebnikUW95} present a family of high-girth bipartite graphs, but their method cannot be adapted to non-bipartite graphs.

\paragraph{Short-cycle removal}
% We briefly note a few works on removing short cycles from existing graphs.

\textcite{Paredes21} gives a polynomial-time algorithm that, given a $d$-regular $(r, \tau)$-graph (that is, such that each node has at most one cycle within $r$~hops, and has at most $\tau$~cycles of length at most~$r$), where $r \leq \frac{2}{3} \log_{d - 1}\left(\frac{n}{\tau}\right) - 5$, outputs a graph with girth~$g \geq r$, while ensuring all eigenvalues remain unchanged except for a bounded factor.
Briefly, the algorithm works by breaking up all short cycles by removing an arbitrary edge in each, and then adding new edges to restore the spectrum, without reintroducing short cycles.
Though this work is the closest to our research question, it does not explicitly investigate the effect stretching has on the convergence speed.

Finally, \textcite{HuEA05} and \textcite{LauTT11} both present what are effectively modifications of the aforementioned work by \textcite{Chandran03} specifically for bipartite graphs.

    \section{Optimal Graph Stretching Problem}\label{sec:optimal-graph-stretching-problem}
% We formalise our optimisation problem and introduce our research questions.

We consider the problem of increasing the girth of a connected graph~$G = (V, E)$ to some~$g \geq 3$ while achieving maximal distributed averaging convergence speed and ensuring that the graph has (almost) no leaves.
Formally, the problem is to find

\begin{equation*}
    \begin{aligned}
        % @formatter:off
        \max_{E' \subseteq V \times V} \quad& \text{convergence speed of\ } H \coloneqq (V, E') \\
        \textrm{such that}             \quad& H\text{\ is a simple connected graph} \\
                                       \quad& \abs{\{ v \in V : \deg(v) < 2 \}} = 0 \\
                                       \quad& \girth(H) \geq g
        % @formatter:on
    \end{aligned}
\end{equation*}

Since this problem is non-linear, it is hard to solve efficiently.
Therefore, we relax our problem definition as follows:
\begin{itemize}
    \item
    Finding the exact convergence speed of a graph is hard.
    Therefore, we settle for a heuristic;
    recall \autoref{subsec:related-work:convergence}.

    \item
    As seen in Moore's bound, there is a difficult-to-control interaction between girth and the number of edges.
    Therefore, we tolerate the presence of some leaves, as long as a best-effort attempt is made.

    % \item
    % We approach the problem in two steps:
    % we opt to divide the problem into smaller problems and solve those in sequence:
    % First, we find a graph that satisfies the constraints, then we minimise the number of leaves, and finally we maximise convergence.
    % In each sub-problem, we ensure that previously attained results are not violated by the current optimisation.
\end{itemize}

Given this relaxed problem formulation, we ask the following research questions:
\begin{itemize}
    \item How does leaf minimisation affect convergence speed?
    \item What is the effect of different stretching methods on convergence speed?
    \item What heuristic achieves maximal convergence speed?
\end{itemize}

We describe our method in \autoref{sec:method} and present our results in \autoref{sec:results}.

    \section{Method}\label{sec:method}
We present our method for answering the questions posed in \autoref{sec:optimal-graph-stretching-problem}.
At a high level, the way we solve the optimal graph stretching problem is to first modify the given graph to satisfy the constraints, and then greedily optimise for the convergence speed heuristic.
More specifically, our approach consists of the following steps:
\begin{itemize}
    \item Generate a graph.
    (\autoref{subsec:method:generate-graphs})
    \item Increase the girth.
    (\autoref{subsec:method:stretch-graphs})
    \item Minimise the number of leaves.
    (\autoref{subsec:method:minimise-leaves})
    \item Optimise graph using a heuristic.
    (\autoref{subsec:method:optimise-convergence})
    \item Run distributed averaging.
    (\autoref{subsec:method:run-distributed-averaging})
\end{itemize}
We repeat this procedure \qty{100}{times} for each combination of parameters.
We provide more details in the subsequent sections.
Source code for the experiments is publicly available~\cite{DekkerEC25bSource}.
We present the results of our method in \autoref{sec:results}.

\subsection{Generate Graphs}\label{subsec:method:generate-graphs}
The accuracy with which heuristics predict convergence speed varies between graph types~\cite{SirocchiB22}.
Therefore, we generate graphs from four families commonly used to model real-world networks.
Each graph is characterised by its number of nodes~$n$ and some family-specific parameters.
For all graphs, we choose $n$ uniformly randomly from the range~$\{25 \ldots 100\}$.
After fixing a set of parameters, we keep generating graphs until a connected graph is found.
We consider the following graph families:
\begin{itemize}
    \item
    $(n, p)$ Erdős--Rényi graphs, where $p$ determines for each possible edge the probability that it is included.
    We choose $p$ uniformly random from the real range $[\ln(n) / n, 1]$, which is the range such that graphs have overwhelming probability of being connected~\cite{ErdosR60}.
    \item
    $(n, k, p)$ Watts--Strogatz graphs, which have small-world properties (i.e.\ high clustering and low distance), which are generated by connecting each node to the previous~$k$ and next~$k$ nodes (creating a ring lattice), and then rewiring each edge with probability~$p$.
    We choose $k$ uniformly random from the integer range $\left[1, \floor(n / 2)\right)$ and $p$ uniformly random from the real range $[0, 1]$, which is the full range of valid parameters.
    \item
    $(n, m)$ Barabási--Albert graphs, which have scale-free properties (i.e.\ asymptotic degree distribution), which are generated by starting with a star topology with $m + 1$~nodes, and then iteratively adding the remaining nodes.
    Each new node is connected to $m$ random existing nodes, with probabilities proportional to nodes' degrees, without replacement.
    We choose $m$ uniformly random from the integer range $[1, n)$, which is the full range of valid parameters.
    \item
    $(n, r)$ geometric graphs, which represent physical networks, and are generated by placing the nodes uniformly random in the unit square, and connecting pairs of nodes within Euclidean distance at most~$r$.
    We choose $r$ uniformly random from the real range $[1.1 \times \sqrt{\frac{\log(n)}{n \pi}}, 1)$, which is the range such that graphs have overwhelming probability of being connected~\cite{Penrose97}.
\end{itemize}

\subsection{Stretch Graphs}\label{subsec:method:stretch-graphs}
To stretch the girth of a graph to threshold~$g$, all cycles with length below $g$ must be removed.
Trivially, it suffices to find all short cycles and remove one edge from each.
However, since cycles may overlap, this naive method may disconnect the graph, and typically removes more edges than necessary.
We remark that, though the underlying application of our work is a distributed protocol, the graph stretching algorithm itself need not be distributed.

In our experiments, we stretch graphs from girth 3 up to and including 10.
Here, girth~3 represents no stretching at all (because every graph has girth at least 3), and girth~10 was chosen because preliminary experiments revealed that very little happens when stretching to even higher girths.

We consider three approaches for stretching a graph to a desired girth.
All three approaches work by repeatedly removing a specific edge until the girth has reached $g$, but differ in how they select that edge:
\begin{itemize}
    \item \textbf{Random:} Remove an edge that is part of a cycle.
    \item \textbf{Least-Cycles:} Remove the edge that is part of the smallest number of shortest cycles.
    \item \textbf{Most-Cycles:} Remove the edge that is part of the largest number of shortest cycles.
\end{itemize}
Each approach considers only those edges that can be removed without disconnecting the graph.
When multiple edges match the criterion, one such edge is chosen randomly.

\begin{remark}
    Note that the third method is expected to remove the most edges.
    We include it nonetheless because the subsequent optimiser in \autoref{subsec:method:optimise-convergence} may benefit from starting with fewer edges.
    \label{remark:method:stretch-graphs:short-cycles-included-anyway}
\end{remark}

\begin{remark}
    Note that the second and third method consider the \enquote{number of shortest cycles}, not the \enquote{number of short cycles}.
    If the graph currently has girth~$g'$, then only cycles with exactly length~$g'$ are counted.
    Eventually, the graph reaches girth~$g' + 1$, and only cycles with exactly length~$g' + 1$ will be counted, and so on until the graph reaches girth~$g$.
    The reason for this is that the \enquote{number of short cycles} quickly becomes infeasibly large, while the \enquote{number of shortest cycles} remains much smaller.
    For example, the complete graph with \qty{25}{nodes} has \qty{2300}{cycles} of length~3, \num{10 626 000}~cycles of length~6, and a frankly immense number of cycles of length~9.
    However, if we first (iteratively) stretch to girth~8, then counting cycles of length~9 becomes feasible again.
    \label{remark:method:stretch-graphs:shortest-not-short-cycles}
\end{remark}

Finding all cycles with length equal to the graph's girth can be done using a simple depth-first search.
We perform this search once at the start, and once again whenever the girth increases.
We store the results in a sparse matrix with a row for each cycle and a column for each edge, similar to an incidence matrix.
(If cycles are hyperedges, then this is the incidence matrix of that hypergraph.)
When an edge is removed, its column is removed from the matrix, and so are all rows that contained that edge.
This way, rows always correspond exactly to eligible cycles, and columns to edges that can be removed without disconnecting the graph.
Finding the edge that is in the largest number of cycles is simply a case of finding the column with the largest number of non-zero values.
Columns can be mapped to edges by keeping track in a map.
% In the worst-case, each edge must be removed, thus querying each edge for each edge.
% Therefore, the worst-case time complexity for stretching is~$\mathcal{O}(g (\abs{V} + \abs{E}) + \abs{E}^2)$, regardless of which variant is chosen.

\subsection{Minimise Leaves}\label{subsec:method:minimise-leaves}
We minimise the number of leaves in the graph without removing nodes and without reducing girth below the threshold~$g$.
We present three methods, which are variations of one algorithm.
We repeat each experiment four times:
once for each method, and once without leaf minimisation.

The high-level algorithm works by iteratively adding edges between pairs of nodes.
To ensure the girth does not sink below~$g$, pairs with distance strictly less than $g - 1$ are excluded.
Initially, the algorithm only connects leaves to other leaves, but when no suitable pairs remain, the algorithm moves on to connect leaves to non-leaves.
The algorithm terminates when no suitable pairs remain.

The three leaf minimisation methods we propose all use the above algorithm but differ in how they choose which pair to connect from the list of candidates:
\begin{itemize}
    \item \textbf{Random:} Connect a random pair of nodes.
    \item \textbf{Closest:} Connect the pair of nodes with the shortest distance.
    \item \textbf{Furthest:} Connect the pair of nodes with the largest distance.
\end{itemize}

This method may fail to remove all leaves in some cases.
For example, when girth is stretched to $g = 4$, this may create a star topology, after which adding an edge will always reduce the girth to $g = 3$.
In this case, our method will not add any edges.
As noted in \autoref{sec:optimal-graph-stretching-problem}, this is acceptable.

\subsection{Optimise Convergence}\label{subsec:method:optimise-convergence}
After minimising the number of leaves and stretching the graph to the desired girth, we optimise the graph's convergence speed for distributed averaging.
We run a greedy algorithm that adds or removes edges until any such change would worsen the convergence speed.
To estimate the convergence speed, we employ a heuristic.
We do not allow the addition or removal of edges that would disconnect the graph, add leaves, or decrease girth below the desired value.

Our choice of heuristics is based on \autoref{subsec:related-work:convergence}:
We choose two graph metrics that are known the correlate well with convergence speed~\cite{SirocchiB22}, and two spectral metrics known to provide bounds on convergence speed~\cite{GhoshB06}.
We repeat each experiment several times, once for each heuristic:
\begin{itemize}
    \item
    \textbf{Eigenratio.}
    Equals $\frac{\lambda_2}{\lambda_n}$.
    Maximised.

    \item
    \textbf{Algebraic connectivity.}
    Equals $\lambda_2$.
    Maximised.

    \item
    \textbf{Closeness centrality.}
    Equals $\sum_{u \in V}\left(\frac{\abs{V} - 1}{\sum_{v \in V} d_{u, v}}\right) / \abs{V}$, given pairwise distances $d$.
    Maximised.

    \item
    \textbf{Global efficiency.}  Equals $\frac{1}{\abs{V} (\abs{V} - 1)} \sum_{u, v \in V, u \neq v} \frac{1}{d_{u, v}}$, given pairwise distances $d$.
    Maximised.
\end{itemize}

\begin{remark}
    The choice for maximisation (rather than minimisation) is based on preliminary results that show that, in our setting, each of these heuristics correlates positively with convergence speed.
\end{remark}

\begin{remark}
    We do not consider clustering coefficient as a metric because, for girth larger than four, the clustering coefficient is always zero by definition.
\end{remark}

We efficiently choose edge removal candidates by finding a cycle basis of the current candidate graph.
This reveals the list of all edges which are in any cycle of any length.
These are exactly the edges that can be removed without disconnecting the graph, since an edge is part of a cycle if and only if the two end nodes have at least two different paths to each other.

We efficiently choose edge addition candidates by finding all pairs of nodes with distance at least $g - 1$.
Adding an edge anywhere else would create a short cycle.

The above operations and heuristics require knowing at each iteration the adjacency matrix, degree matrix, and pairwise distances.
Instead of constantly recalculating these, we calculate these for the initial graph and \enquote{patch} them when an edge is added or removed.
These patches all take constant time, except for patching the pairwise distances when an edge is added, which requires a complete recalculation.

% Overall, time complexity of each greedy iteration consists of calculating the heuristic for each edge candidate, while overhead takes constant time.
% There are $\mathcal{O}(\abs{E}^2)$~candidates, the filtering of which also takes~$\mathcal{O}(\abs{E}^2)$.

\subsection{Run Distributed Averaging}\label{subsec:method:run-distributed-averaging}
We use the asynchronous push-pull model with single-neighbour selection, as described in \autoref{subsec:preliminaries:distributed-averaging}.
At any point in time, given the vector of initial values~$x$ and the vector of intermediate values~$\bar{x}$, we define the error norm as $\frac{\norm{\bar{x} - x}_2}{\norm{x}_2}$.

Each node is assigned an integer value uniformly random from the range~$[0, 50]$.
We continue the protocol until the error norm is less than 0.01.
The convergence time is then the number of rounds taken until convergence is achieved.
For each experiment, to control for randomness, we run 10 instances of distributed averaging, and take the mean convergence time.

The range of starting values does not affect the output;
only the variance does.
Similarly, the exact error norm bound does not qualitatively affect our results.

    \section{Results}\label{sec:results}
We present the results obtained through the method in \autoref{sec:method}.
Firstly, we look at the impact that girth stretching has on convergence speed, without considering leaf removal and optimisation in \autoref{subsec:results:stretching}.
Secondly, we consider the impact of leaf removal in \autoref{subsec:results:leaf-minimisation}.
Finally, we look at the real meat of this work, which is the comparison of various heuristics, specifically when combined with stretching and leaf removal in \autoref{subsec:results:optimisation}.

\subsection{Stretching}\label{subsec:results:stretching}
We look at how stretching affects the graph.
For each stretching method, we look at the number of edges removed, number of leaves created, optimisation heuristics, and convergence time.

\begin{figure}[htb]
    \centering

    \begin{subfigure}[t]{.44\linewidth}
        \centering
        \tikzsetnextfilename{fig:results:stretching:edges:removed}
        \includegraphics[width=\linewidth, height=.75\linewidth]{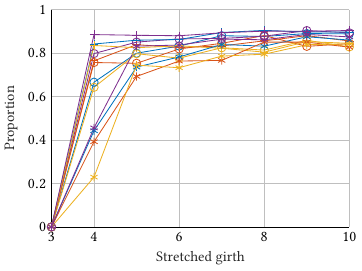}
        \caption{Proportion of edges removed during stretching}
        \label{fig:results:stretching:edges:removed}
    \end{subfigure}%
    \hfil%
    \begin{subfigure}[t]{.44\linewidth}
        \centering
        \tikzsetnextfilename{fig:results:stretching:edges:leaves}
        \includegraphics[width=\linewidth, height=.75\linewidth]{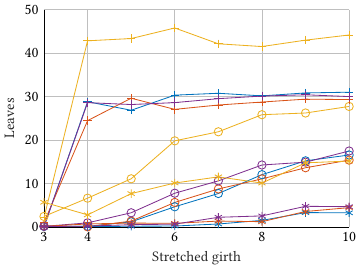}
        \caption{Number of leaves after stretching}
        \label{fig:results:stretching:edges:leaves}
    \end{subfigure}%

    \begin{subfigure}{.8\linewidth}
        \centering
        \tikzsetnextfilename{fig:results:stretching:edges:legend}
        \includegraphics[width=\linewidth]{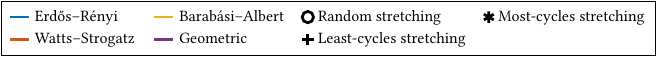}
    \end{subfigure}%

    \caption{Analysis of edges after stretching to a desired girth}
    \label{fig:results:stretching:edges}
\end{figure}

\paragraph{Edges and leaves}
In \autoref{fig:results:stretching:edges:removed}, we show the proportion of edges removed by stretching for each combination of graph family and stretching method.
Note that a girth of three implies that no stretching has taken place.
Though the proportion quickly increases for all experiments, it also immediately flattens out.
This shows that, at least in these graph families, removing all short cycles is typically sufficient to remove the majority of longer cycles (see \autoref{remark:method:stretch-graphs:shortest-not-short-cycles}).
As expected, the most-cycles stretching method removes the smallest proportion of edges, followed by random stretched, and then least-cycles stretching.
Watts--Strogatz graphs and Barabási--Albert graphs require removing the smallest proportion of edges;
their being highly clustered means that most cycles are centred around just a few edges, which are quickly removed.
However, as girth increases, differences between graph types and stretching methods diminish into the negligible.

In \autoref{fig:results:stretching:edges:leaves}, we show the number of leaves in stretched graphs.
All graphs have (nearly) no leaves at girth~3, which is before any stretching takes place.
The number of leaves quickly goes up when the graph is stretched, with major differences between graph types and stretching methods.
Among graph types, we observe that Barabási--Albert graphs have significantly more leaves than all other graph types regardless of which stretching method is used.
This is because these graphs have many low-degree nodes, so the removal of any edge is likely to create a new leaf.
When we compare stretching methods, we see that, regardless of graph type, most-cycles stretching creates very few new leaves even when stretching to girth~10, random
stretching performs approximately three times as badly, and least-cycles stretching shoots up so quickly that it hits a ceiling because the stretched graph is (nearly) a tree.

\begin{figure}[p]
    \centering

    \begin{subfigure}[t]{\linewidth}
        \centering
        \tikzsetnextfilename{fig:results:stretching:convergence-heuristic:grid}
        \includegraphics[width=\linewidth, height=.75\linewidth]{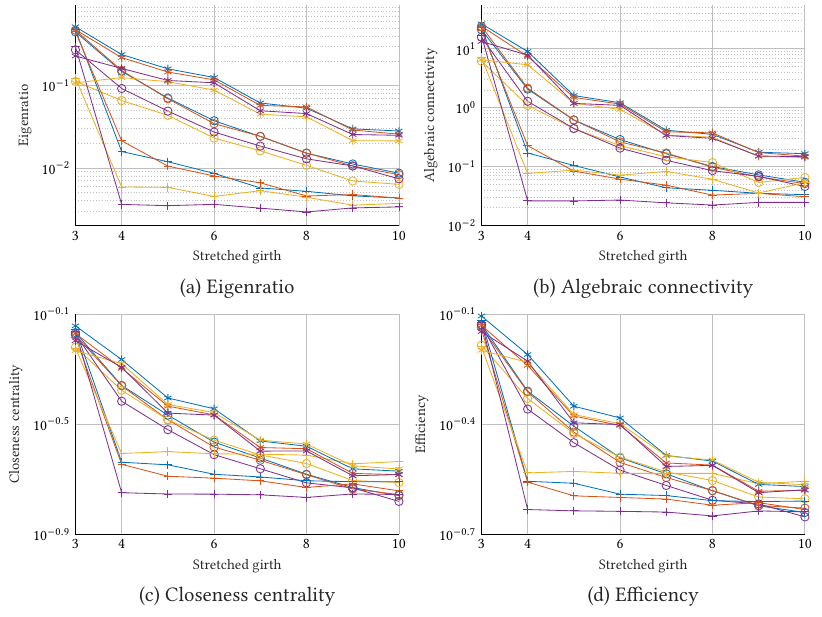}
        \label{fig:results:stretching:convergence-heuristic:grid}
    \end{subfigure}%
    \vspace{-4mm}%

    \begin{subfigure}{\linewidth}
        \centering
        \tikzsetnextfilename{fig:results:stretching:convergence-heuristic:legend}
        \includegraphics{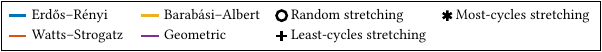}
    \end{subfigure}%
    \vspace{-2mm}%

    \caption{Convergence heuristics after stretching to a desired girth}
    \label{fig:results:stretching:convergence-heuristic}

    \bigskip\bigskip

    \begin{subfigure}[c]{.6\linewidth}
        \centering
        \tikzsetnextfilename{fig:results:stretching:convergence-time:figure}
        \includegraphics[width=\linewidth, height=.75\linewidth]{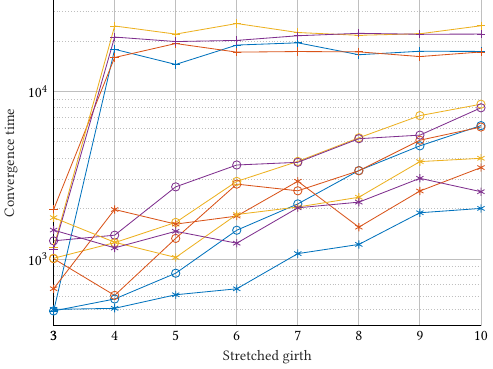}
        \label{fig:results:stretching:convergence-time:figure}
    \end{subfigure}%
    \begin{subfigure}[c]{.3\linewidth}
        % 1.6 = (.8 * .6) / .3
        \centering
        \raisebox{\dimeval{1.6\linewidth-\height}}{%
            \tikzsetnextfilename{fig:results:legend-with-colours-for-layouts-and-markers-for-stretching-methods:vertical}
            \includegraphics{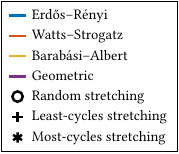}
        }
        \label{fig:results:stretching:convergence-time:legend}
    \end{subfigure}%
    \vspace{-4mm}%

    \caption{Convergence time after stretching}
    \label{fig:results:stretching:convergence-time}
\end{figure}

\paragraph{Convergence heuristics}
In \autoref{fig:results:stretching:convergence-heuristic}, we show the convergence time heuristics for stretched graphs.
In all cases, higher is better.
The four heuristics behave quite similarly, predicting worse convergence time as girth increases, but predicted performance flattens out at higher girths.
Across graph types, all heuristics predict that Barabási--Albert graphs and geometric graphs perform worse when stretched to low girths, but joins up with the rest once stretched to girth~10.
Across stretching methods, least-cycles stretching typically drops down immediately before flooring out, while random stretching and most-cycles stretching approach this floor gradually with increased girth, with the latter keeping higher predicted convergence times.

We note that the most-cycles stretching method exhibits a \enquote{sawtooth} pattern, where heuristics drop harder at odd values than at even values.
When we inspect cycle counts in individual graphs, we find that stretching to an even girth typically also removes a disproportionate amount of odd-length cycles, even those longer than the desired girth.
For example, after stretching to girth~4 with the most-cycles method, the resulting graphs often end up having fewer length-7 cycles than length-6 cycles, even though this is not true for any of the unstretched graphs.
This holds even if we use a variant of the most-cycles stretching method that counts \emph{all} cycles, not just the shortest ones (see \autoref{remark:method:stretch-graphs:shortest-not-short-cycles}).
This effect is most pronounced in Barabási--Albert graphs.

\paragraph{Convergence time}
In \autoref{fig:results:stretching:convergence-time}, we show the empirical convergence time for stretched graphs.
Lower is better.
It is immediately clear that least-cycles stretching performs terribly, presenting a fourfold increase compared to random stretching, and a sevenfold increase in convergence time compared to most-cycles stretching.
We see from \autoref{fig:results:stretching:convergence-heuristic} that the heuristics are decent predictors of convergence time, though the predicted divide between graph types is not present in the empirical measurements.
We note, however, that an even better predictor of performance is the number of leaves removed (see \autoref{fig:results:stretching:edges:leaves}), or rather, the number of nodes removed.

We conclude that convergence time is seriously impacted by stretching, but that this is not due to cycle removal per se, but due to the removal of many edges.
Therefore, most-cycles stretching is the optimal method, despite its sawtooth behaviour.

\subsection{Leaf Minimisation}\label{subsec:results:leaf-minimisation}
We look at how effectively leaf minimisation removes leaves, and at its effect on convergence time.

\begin{figure}[p]
    \centering

    \begin{subfigure}[t]{.8\linewidth}
        \centering
        \tikzsetnextfilename{fig:results:leaf-minimisation:leaves-remaining:grid}
        \includegraphics[width=\linewidth, height=.75\linewidth]{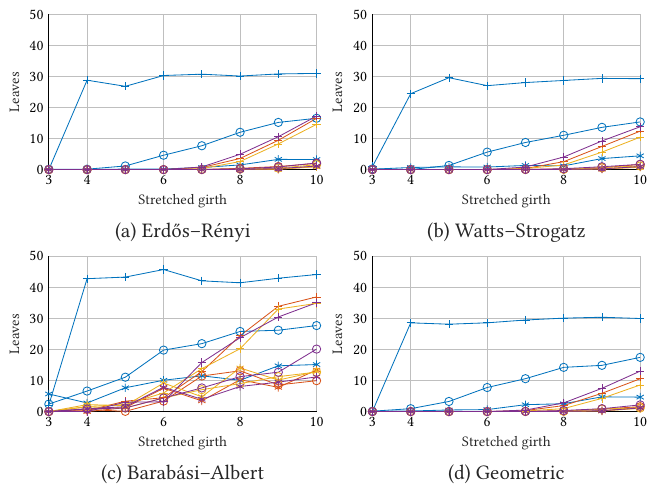}
        \label{fig:results:leaf-minimisation:leaves-remaining:grid}
    \end{subfigure}%
    \vspace{-4mm}%

    \begin{subfigure}{\linewidth}
        \centering
        \tikzsetnextfilename{fig:results:leaf-minimisation:leaves-remaining:legend}
        \includegraphics{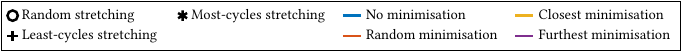}
    \end{subfigure}%
    \vspace{-2mm}%

    \caption{Number of leaves remaining after leaf minimisation}
    \label{fig:results:leaf-minimisation:leaves-remaining}

    \bigskip

    \begin{subfigure}[t]{.8\linewidth}
        \centering
        \tikzsetnextfilename{fig:results:leaf-minimisation:edges-added:grid}
        \includegraphics[width=\linewidth, height=.75\linewidth]{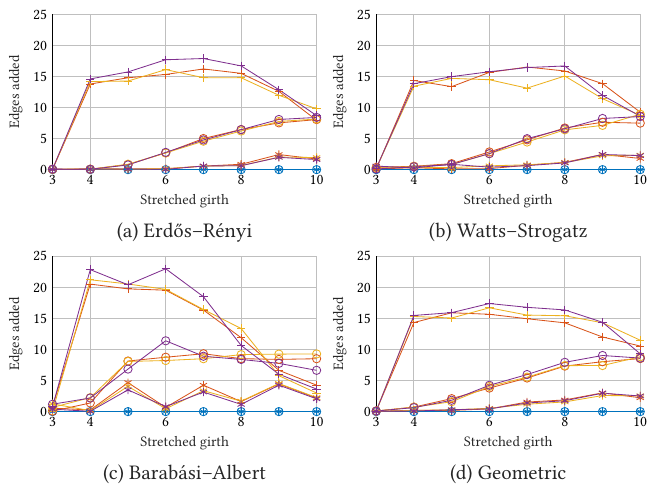}
        \label{fig:results:leaf-minimisation:edges-added:grid}
    \end{subfigure}%
    \vspace{-4mm}%

    \begin{subfigure}{\linewidth}
        \centering
        \tikzsetnextfilename{fig:results:leaf-minimisation:edges-added:legend}
        \includegraphics{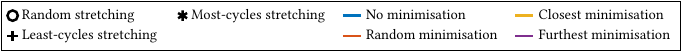}
    \end{subfigure}%
    \vspace{-2mm}%

    \caption{Number of edges added during leaf minimisation}
    \label{fig:results:leaf-minimisation:edges-added}
\end{figure}

\paragraph{Leaves and edges}
In \autoref{fig:results:leaf-minimisation:leaves-remaining}, we show the number of leaves that remain after leaf minimisation.
(Note that, unlike previous graphs, colours indicate leaf minimisation method, not graph family.)
The lines representing no leaf minimisation correspond exactly to \autoref{fig:results:stretching:edges:leaves}.
When we compare stretching methods, we see that least-cycles stretching creates the largest number of leaves, followed by random stretching, and then most-cycles stretching, though the latter two are close.
When we compare leaf minimisation methods, we see only small differences, with closest leaf minimisation most effectively eliminating leaves, followed by random leaf minimisation, and finally furthest leaf minimisation.
There are no significant differences between graph types.

In \autoref{fig:results:leaf-minimisation:edges-added}, we show the number of edges added by leaf minimisation.
Recall that our minimisation method starts by connecting leaves to each other before connecting leaves to non-leaves, and thus the number of edges added is not necessarily linear in the number of leaves eliminated.
The lines for most-cycles stretching and random stretching are similar to their counterparts in \autoref{fig:results:leaf-minimisation:leaves-remaining}, whereas the least-cycles stretching line goes down when girth goes up.
The latter result is visible in \autoref{fig:results:leaf-minimisation:leaves-remaining}:
The number of leaves before minimisation hits a ceiling and stays the same, while the number of leaves after minimisation increases.
Thus, fewer leaves have been eliminated, and therefore fewer edges must have been added.
Overall, this implies that the graph's diameter (the length of the longest shortest path) resulting from least-cycles stretching is too small to allow leaf minimisation without reducing girth.

\paragraph{Convergence time}
In \autoref{fig:results:leaf-minimisation:convergence-time}, we show the convergence time after leaf minimisation.
There are no significant differences between leaf minimisation methods.
Though the sawtooth pattern with most-cycles stretching complicates the graphs, it is clear that leaf minimisation improves convergence time for all stretching methods, especially least-cycles stretching.
However, we argue that it is not the leaf minimisation itself that improves the convergence, but simply the fact that \emph{any} edges are added to the graph.
This is apparent from the lack of similarity to \autoref{fig:results:leaf-minimisation:leaves-remaining} and \autoref{fig:results:leaf-minimisation:edges-added}.
We conclude that leaf minimisation is neither detrimental nor beneficial to performance.

\subsection{Optimisation}\label{subsec:results:optimisation}
Finally, we look at the effect of optimising convergence time with heuristics.

\begin{figure}[p]
    \centering

    \begin{subfigure}[t]{.8\linewidth}
        \centering
        \tikzsetnextfilename{fig:results:leaf-minimisation:convergence-time:grid}
        \includegraphics[width=\linewidth, height=.75\linewidth]{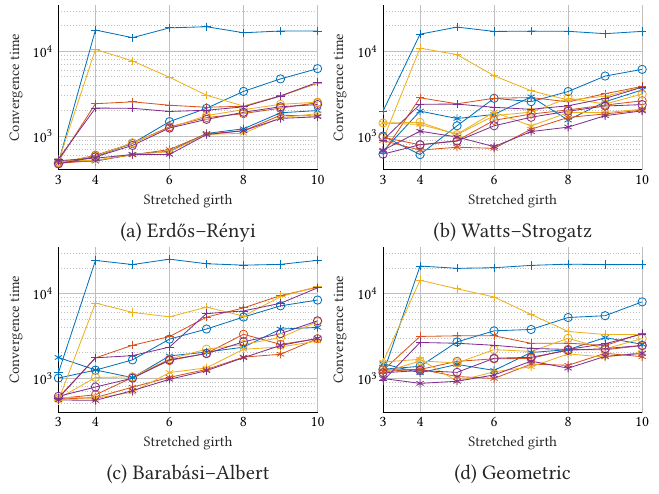}
        \label{fig:results:leaf-minimisation:convergence-time:grid}
    \end{subfigure}%
    \vspace{-4mm}%

    \begin{subfigure}{\linewidth}
        \centering
        \tikzsetnextfilename{fig:results:leaf-minimisation:convergence-time:legend}
        \includegraphics{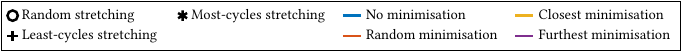}
    \end{subfigure}%
    \vspace{-2mm}%

    \caption{Convergence time after leaf minimisation}
    \label{fig:results:leaf-minimisation:convergence-time}
    \bigskip

    \begin{subfigure}[t]{.8\linewidth}
        \centering
        \tikzsetnextfilename{fig:results:optimisation:edges-changed:grid}
        \includegraphics[width=\linewidth, height=.75\linewidth]{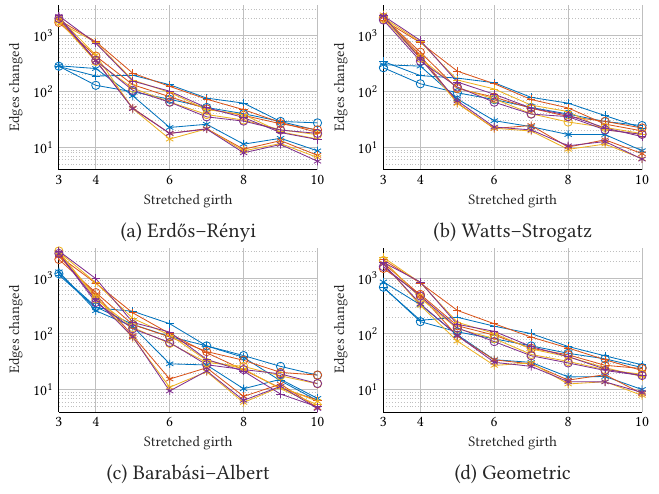}
        \label{fig:results:optimisation:edges-changed:grid}
    \end{subfigure}%
    \vspace{-4mm}%

    \begin{subfigure}{\linewidth}
        \centering
        \tikzsetnextfilename{fig:results:optimisation:edges-changed:legend}
        \includegraphics{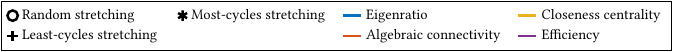}
    \end{subfigure}%
    \vspace{-2mm}%

    \caption{Number of edges added or removed during optimisation}
    \label{fig:results:optimisation:edges-changed}
\end{figure}

\paragraph{Number of edges}
In \autoref{fig:results:optimisation:edges-changed}, we show the number of edges added or removed during optimisation, without considering leaf optimisation.
Intuitively, this is a measure of how many steps stretched graphs are removed from the optimum.
On average, graphs have 238~edges before optimisation and 380~edges after optimisation, with significantly more edges added than removed.
However, the number of changes decreases as girth increases.
Though greedy algorithms may get stuck in local optima, additional experiments using simulated annealing based on the method by~\textcite{MahdiA09} show that even search algorithms without this drawback require a decreasing number of changes to the edge set.
The downwards trend thus appears to be inherent to the optimal graph stretching problem itself.

When we compare graph types, we see that they differ only in scale, with Barabási--Albert graphs requiring the most changes.
In all four graph types, stretched graphs require the fewest changes after most-cycles stretching, followed by random stretching, and then least-cycles stretching.
The only exception is low-girth graphs optimised by eigenratio, where all stretching methods perform similarly.

\begin{figure}[htb]
    \centering

    \begin{subfigure}{\linewidth}
        \centering
        \tikzsetnextfilename{fig:results:optimisation:convergence-time}
        \includegraphics[width=\linewidth]{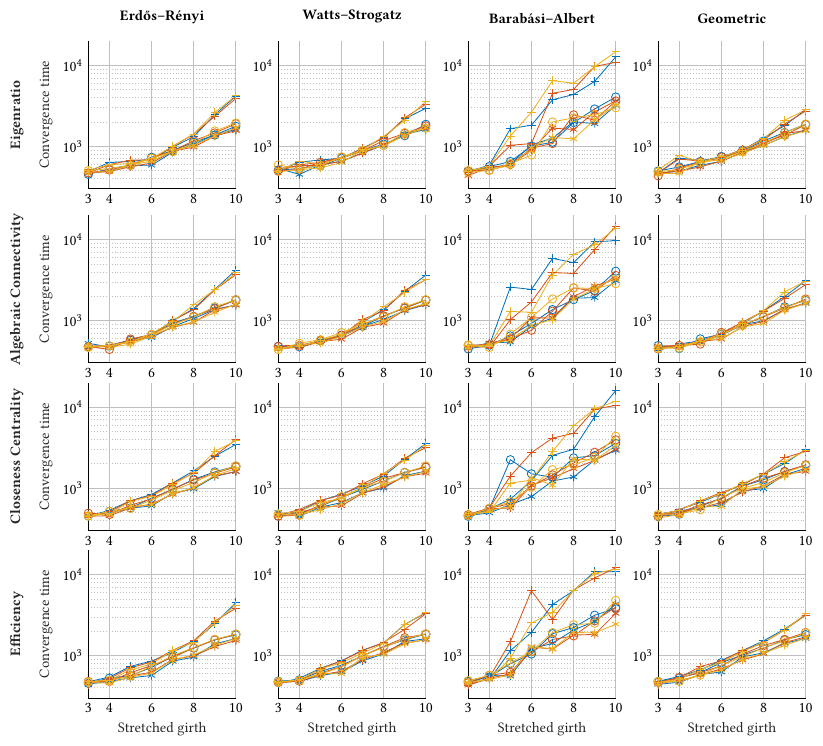}
        \label{fig:results:optimisation:convergence-time}
    \end{subfigure}%

    \begin{subfigure}{\linewidth}
        \centering
        \tikzsetnextfilename{fig:results:optimisation:convergence:legend}
        \includegraphics[width=\linewidth]{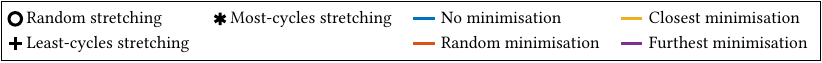}
    \end{subfigure}%

    \caption{Convergence time after heurisical optimisation, with columns indicating the graph type, and rows indicating the heuristic that was optimised}
    \label{fig:results:optimisation:convergence}
\end{figure}

\paragraph{Convergence time}
In \autoref{fig:results:optimisation:convergence}, we show the effect of heuristic optimisation on convergence time per graph family.
(Note the different y-axis scale per column.)
All sixteen graphs have many similarities.
When we compare stretching methods, most-cycles stretching and random stretching achieve the lowest convergence time, followed by least-cycles stretching, defeating the hypothesis that the optimiser may benefit from fewer edges being removed.
When we consider leaf minimisation, we see that there is little difference between the various methods, and confirm that leaf minimisation by itself is not responsible for improved convergence time.
When we compare heuristics, we also do not see a clear winner.
Though graphs stretched with the least-cycles method appear to benefit from choosing the right heuristic for the graph type, differences are much smaller for the other stretching methods.
Finally, several figures, especially those describing Barabási--Albert graphs, contain the aforementioned sawtooth pattern.

    \section{Conclusion}\label{sec:conclusion}
We investigated the relation between a graph's girth and the convergence time of distributed averaging.
We introduced the \emph{optimal graph stretching problem}, which is the task of increasing the girth of a graph while keeping the convergence time and number of leaves minimal, and the graph connected.
We proposed and implemented a sequence of algorithms to solve this problem, which we applied to hundreds of thousands of graphs, after which we measured the results.

We find that stretching the girth of a graph increases convergence time proportional to the number of edges removed.
Consequently, stretching by iteratively removing the edge that is simultaneously in the largest number of cycles results in the smallest convergence time cost.
Furthermore, convergence time can be recuperated using a greedy algorithm to add edges without decreasing girth.
Finally, minimising the number of leaves does not affect convergence time.

We note a few possible avenues for future work.
Firstly, the aforementioned stretching method creates a sawtooth pattern in the distribution of cycle lengths, which may be of independent interest.
Secondly, the studied heuristics correlate worse with convergence time than in related work;
we postulate that our results may be improved by developing high-girth-specific heuristics.
Finally, our solution to the optimal graph stretching problem requires global knowledge of the graph, but for ad-hoc networks it may be useful to create a distributed solution.

    \hbadness=10000\hfuzz=10000pt\printbibliography
\end{document}